\documentclass[runningheads]{llncs}
\usepackage[T1]{fontenc}
\usepackage{graphicx}
\usepackage{cite}  
\usepackage{hyperref}
\begin{document}
\title{A shared neural encoding model for the prediction of subject-specific fMRI response} 
\titlerunning{Prediction of subject-specific fMRI response}
%Inter-subject knowledge transfer in neural encoding of naturalistic stimuli}

%\title{Individualized prediction of fMRI response during naturalistic stimulation}
\author{Anonymous Authors
}
%index{Khosla, Meenakshi}

\author{Meenakshi Khosla\inst{1} \and
Gia H. Ngo\inst{1} \and
Keith Jamison\inst{2} \and
Amy Kuceyeski\inst{2,3} \and
Mert R. Sabuncu\inst{1,2,4}}
\authorrunning{Khosla et al.}

% \authorrunning{Anonymous et al.}
% \institute{Anonymous Institution}
%
\institute{1. School of Electrical \& Computer Engineering, Cornell University\\
2. Radiology, Weill Cornell Medical College \\
3. Brain and Mind Research Institute, Weill Cornell Medical College \\
4.  Nancy E. \& Peter C. Meinig School of Biomedical Engineering, Cornell University}

\maketitle              % typeset the header of the contribution
\begin{abstract}
The increasing popularity of naturalistic paradigms in fMRI (such as movie watching) demands novel strategies for multi-subject data analysis, such as use of neural encoding models. 
In the present study, we propose a shared convolutional neural encoding method that accounts for individual-level differences. 
Our method leverages multi-subject data to improve the prediction of subject-specific responses evoked by visual or auditory stimuli. 
We showcase our approach on high-resolution 7T fMRI data from the Human Connectome Project movie-watching protocol and demonstrate significant improvement over single-subject encoding models. 
We further demonstrate the ability of the shared encoding model to successfully capture meaningful individual differences in response to traditional task-based facial and scenes stimuli. Taken together, our findings suggest that inter-subject knowledge transfer can be beneficial to subject-specific predictive models.\footnote{Our code is available at \url{https://github.com/mk2299/SharedEncoding_MICCAI}.}  
%Our code is freely available at [Add anonymous link]. 
%suggesting the potential of naturalistic stimuli as a one-paradigm-fits-all solution to brain mapping and move beyond the carefully-constructed task-based approaches that often use artificial stimuli. 
%The abstract should briefly summarize the contents of the paper in
%15--250 words.

%\keywords{fMRI  \and encoding \and audition \and vision \and naturalistic stimuli}
\end{abstract}
\section{Introduction}

Naturalistic imaging paradigms, such as movies and stories, emulate the diversity and complexity of real-life sensory experiences, thereby opening a novel window into the brain. 
The last decade has seen an increased foothold of naturalistic paradigms in cognitive neuroimaging, fueled by the remarkable discovery of inter-subject synchrony during naturalistic viewing~\cite{pmid15016991}. 
Naturalistic stimuli also demonstrate increased test-retest reliability and more active subject engagement in comparison to alternate paradigms such as resting-state fMRI~\cite{pmid31257145}. 
Furthermore, experiments have shown that naturalistic stimuli can induce stronger neural response than task-based stimuli~\cite{Schultz2009}, suggesting that the brain is intrinsically more attuned to the former. Taken together, these benefits suggest an exciting future for naturalistic stimulation protocols in fMRI. 
% For example, facial stimuli in natural motion were shown to evoke higher neural activations in face-specialized regions than artificial, static faces

With large-scale compilation of multi-subject neural data through open-source initiatives such as the Human Connectome Project (HCP) \cite{hcp}, the development of approaches that can handle this enormous data is becoming imperative. Two approaches, namely inter-subject correlation (ISC) analysis \cite{pmid15016991, pmid20004608} and shared response model (SRM) \cite{Chen2015ARF}, have dominated the analysis of multi-subject fMRI data under naturalistic conditions. The former approach exploits similarity in activation patterns across subjects to isolate stimulus-induced processing. The latter technique, SRM, decomposes neural activity into a shared response component and subject-specific spatial bases, and has been used for inter-subject knowledge transfer through functional alignment. While simple and efficient, both these approaches rely on a common time-locked stimulus across subjects and cannot, by design, model responses to completely unseen stimuli. On the other hand, predictive modelling of neural activity through encoding models is based upon generalization to arbitrary stimuli and can thus offer more holistic descriptions of sensory processing in an individual \cite{pmid30513462}. 
%based upon the notion that shared activity must be driven by processing of the common stimulus
%Building an accurate model of brain function under naturalistic stimuli can help in discovering novel relationships between high-level stimulus descriptions and evoked brain response in an individual within a general-purpose hypothesis-free framework.

Neural encoding models map stimuli to fine-grained voxel-level response patterns via complex feature transformations. Previously, neural encoding models have yielded several novel insights into the functional organization of auditory and visual cortices \cite{Kell2018a, pmid26157000, pmid24812127, Wen2018a}. Encoding models encapsulating different hypothesis about neural information processing can be pitted against each other to shed new light on how information is represented in the brain. In this manner, neural encoding models have been largely used for making group-level inferences. The potential to extract meaningful individual differences from naturalistic paradigms remains largely untapped. Understanding inter-subject variability in behavior-to-brain representations is of key interest to neuroscience and can potentially even help identify atypical response patterns \cite{pmid27138646}. Modelling individual brain function in response to naturalistic stimuli is one step in this direction; however, building accurate individual-level models of brain function often requires large amounts of data per subject for good generalization. The problem is further exacerbated by the variability in anatomy and functional topographies across individuals, making inter-subject knowledge transfer difficult. There is limited work in leveraging multi-subject data for more robust and accurate individualized neural encoding. To our knowledge, this problem has been studied only in the context of natural vision with a handful subjects using a Bayesian framework \cite{Wen2018b}. Further, the proposed method in \cite{Wen2018b} transfers  knowledge from one subject's encoding model into another through a two-stage procedure and does not allow simultaneous optimization of encoding models across multiple subjects.

%While existing methods have focused on capturing the common response, recently there is growing interest in characterizing individual differences with naturalistic neuroimaging \cite{pmid31257145}. 
\begin{figure*}[tbhp]
\begin{center}
\includegraphics[width=\textwidth]{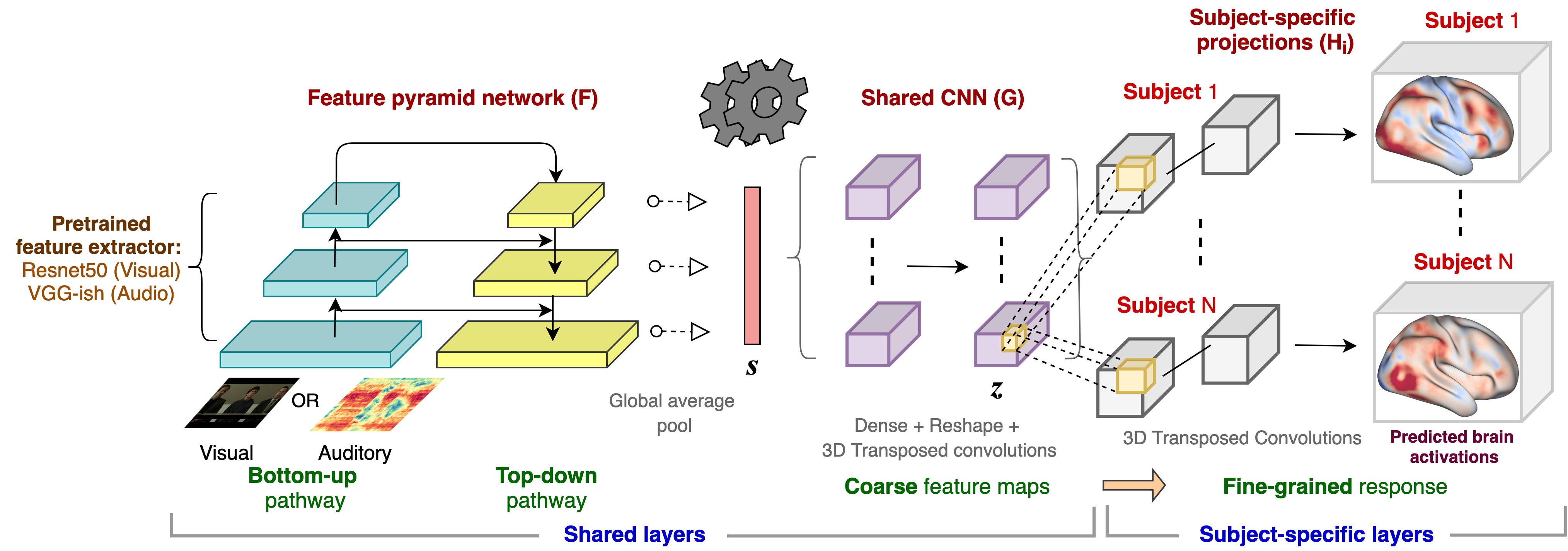}
%\caption{\textbf{Proposed approach}: Hierarchical features are extracted from pre-trained image/sound recognition networks, reshaped into coarse 3D feature maps, and transformed into increasingly fine-grained maps using learnable convolutions. Coarse transformation layers are shared across subjects while deeper convolutional layers close to predicted response are subject-specific.}
\caption{\textbf{Proposed approach:} Feature pyramid networks are used to extract hierarchical features from pre-trained image/sound recognition networks. Dense features are reshaped into coarse 3D feature maps, which are mapped into increasingly fine-grained maps using convolutions. Coarse feature transformation layers are shared across subjects while deeper convolutional layers close to predicted response are subject-specific.}
\label{fig:schematic}
\end{center}
\end{figure*}

In this paper, we attempt to fill this gap; to this effect, we propose a deep-learning based framework to build more powerful individual-level encoding models by leveraging multi-subject data. Recent studies have revealed that coarse-grained response topographies are highly similar across subjects, suggesting that individual idiosyncrasies manifest in more fine-grained response patterns~\cite{Guclu2017, Chen2015ARF}. This hints to the idea that encoding models could share representational spaces across subjects to overcome the challenges imposed by a limited quantity of per-subject data. We exploit this intuition to develop a neural encoding model with a common backbone architecture for capturing shared response and subject-specific projections that account for individual response biases, as demonstrated in Figure \ref{fig:schematic}.  
Our proposed approach has several merits: (i) It allows us to combine data from multiple subjects watching same or different movies to build a global model of the brain. At the same time, it can capture meaningful individual-level deviations from the global model which can potentially be related to individual-specific traits. (ii) It is amenable to incremental learning with diverse, varying stimuli across seen or novel subjects with less constraints on data collection from single subjects. (iii) It poses minimal memory overhead with additional subjects and can thus handle fMRI datasets with a large number of subjects. 
%(iv) It paves the way for diverse sampling of naturalistic stimuli across participants with less constraints on data collection from single subjects 
%(v) It allows joint end-to-end optimization of the feature and response models. 

%Shared spaces can also force the model to extract generalizable representations from each stimulus, thereby granting more validity to the findings of a study. 

\section{Methodology}

Our proposed methodology is illustrated in Figure \ref{fig:schematic}. Neural encoding models comprise two components: (a) a feature extractor, which pulls out relevant features from raw images or audio waveforms and (b) a response model, which maps these stimuli features into brain responses. In contrast to existing works that employ a linear response model~\cite{Wen2018a, pmid26157000}, we propose a CNN-based response model where the coarse 3D feature maps are shared across subjects and fine-grained feature maps are individual-specific. Previous studies have reported a cortical processing hierarchy where low-level features from early layers of a CNN-based feature extractor best predict responses in early sensory areas while semantically-rich deeper layers best predict higher sensory regions \cite{Kell2018a, pmid26157000}. To account for this effect, we employ a hierarchical feature extractor based on feature pyramid networks \cite{Lin2016FeaturePN} that combines features from early, intermediate and later layers simultaneously. The output of the feature extractor is fed into the convolutional response model to predict the evoked fMRI activation. This enables us to train both components of the network simultaneously in an end-to-end fashion. 

Formally, let $\mathcal{D} = \{\textbf{X}_i, \textbf{Y}_i\}_{i=1}^N$ denote the training data pairs for N subjects, where \textbf{X}$_i$ denotes the stimuli presented to subject $i$ and $\textbf{Y}_i$ denotes the corresponding fMRI measurements. We represent \textbf{X}$_i$ as RGB images or grayscale spectrograms for the visual and auditory models, respectively. The feature model maps the 2D input into a vector representation \textbf{s} and is parameterized using a deep neural network \textbf{F}(\textbf{X}$_i; \mathrm{\phi})$ that is common across subjects. In our experiments, this model is a feature pyramid network built upon pre-trained recognition networks as DNNs optimized for image or sound recognition tasks have proven to provide powerful feature representations for encoding brain response. We define a differentiable function \textbf{G}(\textbf{s}$; \mathrm{\theta})$ that maps the features into a shared latent volumetric space \textbf{z}, whose first 3 axes represent the 3D voxel space and the last axis captures the latent dimensionality. The predicted response for each subject is then defined using subject-specific differentiable functions {\textbf{H}$_i$}(\textbf{z}$; \psi_i)$ that project the coarse feature maps \textbf{z} into an individualized brain response. We represent \textbf{G} and \textbf{H}$_i$'s using convolutional neural networks to have a sufficiently expressive model. Thus, $\theta $ and $\{\psi_i\}$ represent a mix of convolutional kernels or dense weight matrices.  The number of shared parameters, $|\theta| + |\phi|$ is kept much greater than the cardinality of subject-specific parameters $|\psi_i|$ to accurately estimate the shared latent space. All parameters $\{\phi, \theta, \psi_i\}$ are trained jointly to minimize the \textit{mean squared error} between the predicted and true response. The proposed method allows us to propagate errors through the shared network even if the subjects are not exposed to common stimuli since we can always backpropagate errors for subjects independently within each batch. Furthermore, using individualized layers to account for subject-specific biases enables the model to weigh gradients coming from losses of each subject differently according to their signal-to-noise ratio. This makes the model less susceptible to noisy measurements when responses for the same stimuli are available from multiple subjects.  
 
\subsection{Implementation details}
We employ pre-trained Resnet-50 \cite{He2015DeepRL} and VGG-ish \cite{Hershey2016CNNAF} architectures in the bottom-up path of Figure \ref{fig:schematic} to extract multi-scale features from images and audio spectrograms, respectively. The base architectures were selected because pre-trained weights of these networks optimized for classification on large datasets, namely Imagenet\cite{Deng2009ImageNetAL} and Youtube-8M\cite{AbuElHaija2016YouTube8MAL}, were publically available. 
%In Resnet50, we keep all layers of the pre-trained network except the fully connected layers to handle images of higher dimensionality (720x1024x3 in our case) whereas for the VGG network, we use all weights, including the dense layers, before the final classification layer of the pre-trained network. 
For Resnet-50, we use activations of the last residual block of each stage, namely, {\small \fontfamily{qbk}\selectfont res2, res3, res4} and {\small \fontfamily{qbk}\selectfont res5} (notation from \cite{Detectron2018}) to construct our stimulus descriptions \textbf{s}. From the VGG network, we use the activations of each convolutional block, namely, {\small \fontfamily{qbk}\selectfont conv2, conv3, conv4} and the penultimate dense layer {\small \fontfamily{qbk}\selectfont fc2}\cite{vggish}. The first three set of activations are refined through a top-down path to enhance their semantic content, while the last activation is concatenated into \textbf{s} directly (res4 activations are vectorized using global average pool). The top-down path comprises three feature maps at different resolutions with an up-sampling factor of 2 successively from the deepest layer of the bottom-up path. Each such feature map comprising 256 channels is merged with the corresponding feature map in the bottom-up path (reduced to 256 channels by 1x1 convolutions) by element-wise addition.  Subsequently, the feature map at each resolution is collapsed into a 256 dimensional feature vector through a global average pool operation and concatenated into \textbf{s}. The aggregated features are then passed onto a shared CNN (denoted \textbf{G} above) comprising the following feedforward computation: a fully connected layer to map the features into a vector space which is reshaped into a 1024-channel cuboid of size {\small \fontfamily{qbk}\selectfont 6x7x6} followed by two {\small \fontfamily{qbk}\selectfont 3x3x3} transposed convolutions (conv.T) with a stride of 2 to up-sample the latter and obtain \textbf{z}.  Each convolution reduces the channel count by half, thereby, resulting in a shared latent \textbf{z} that is a 256 channel cuboid of size {\small \fontfamily{qbk}\selectfont 27x31x27x256}. Subject-specific functions {\textbf{H}$_i$}'s are parameterized as a cascade of two {\small \fontfamily{qbk}\selectfont 3x3x3} conv.T operations (stride 2) with output dimensions 128 and 1 respectively. It is important to emphasize that these operations constitute much fewer parameters, thereby favoring the estimation of a shared truth. As we demonstrate empirically, a shared space allows much better generalization. At the same time, we find that even the limited subject-specific parameters can adequately capture meaningful individual differences. All parameters were optimized using Adam\cite{Kingma2014AdamAM} with a learning rate of 1e-4. Auditory and visual models were trained for 25 and 50 epochs respectively with unit batch size. Validation curves were monitored to ensure convergence. 
%\footnote{Pre-trained tensorflow/keras models were available at \url{https://keras.io/applications} \url{https://github.com/tensorflow/models/tree/master/research/audioset/vggish} respectively}
%The bottom-up path was initialized with these pre-trained weights and the bottom-up path was trained from scratch. 

%\section{Experiments}
\subsection{Data and Preprocessing}
We study 7T fMRI data (TR = 1s) from a randomly selected sample of N=10 subjects from HCP movie-watching protocol \cite{hcp, hcp2}. The dataset comprises 4 audiovisual movies, each $\sim$15 mins long. Preprocessing protocols are described in detail in \cite{pmid27894889, hcp2}. For our experiments, we utilize the 1.6mm MNI-registered volumetric images of size {\small \fontfamily{qbk}\selectfont 113 x 136 x 113} per TR. We compute log-mel spectrograms using same parameters as \cite{Hershey2016CNNAF} over every 1 second of audio waveform to obtain a 2D image-like input for the VGG audio feature extractor. We extract the last frame of every second of the video to present to the image recognition network for visual features. We estimate a hemodynamic delay of $4$ $sec$ using regression based encoding models, as the response latency that yields highest encoding performance. Thus, all proposed and baseline models are trained to use the above stimuli to predict the fMRI response 4 seconds $\textit{after}$ the corresponding stimulus presentation. We train and validate our models on three movies using a $~$9:1 train-val split and leave the fourth movie for independent testing. This yields 2000 training, 265 validation and 699 test stimulus-response pairs per subject.  
%The movies represent a diverse collection, ranging from short snippets of Hollywood movies to independent vimeo clips.
%We note that is a fairly high spatial and temporal resolution compared to existing studies. 

\subsection{Baselines}
%We compare our approach against three other baselines. 
\begin{itemize}
\item Linear response model (individual subject): Here, we train independent models for each subject using linear response models. We note that, thus far, this is the dominant approach to neural encoding. To enable a fair comparison, we extract hierarchical features of the same dimensionality as the proposed model to present to the linear regressor. The only difference here is the lack of a top-down pathway (since it is not pre-trained), which prevents the refinement of coarse feature maps before aggregation. We apply $l_2$ regularization on the regression coefficients and adjust the optimal strength of this penalty through cross-validation using log-spaced values in $\{1\mathrm{e}{-10}, 1\mathrm{e}{10}\}$. We report the performance of the best model as `Individual (Linear)'.  
\item CNN response model (individual subject): 
Here, we employ the same architecture as the proposed model but with only one branch of subject-specific layers. 
We train this network independently for each subject without weight sharing and denote its performance as `Individual model (CNN)'.  
\item Shared model (mean): Here, we employ the proposed model after training but instead of computing predictions using the same subject's learned weights, we compute $N$ predictions from all subject-specific branches. We compute the mean performance obtained by correlating each of these predictions with the ground truth response of a subject and denote this as `Shared (mean)'. 

\end{itemize}
%using features from all layers yielded better performance than using features from any single layer alone, so we used all the features for prediction. 

\subsection{Performance evaluation}
We measure performance on the test movie by computing the \textit{Pearson's correlation coefficient} between the predicted and measured fMRI response at each voxel. Since different subjects have a different signal-to-noise ratio, we normalize each voxel's correlation by the subject's noise ceiling for that voxel. We compute the subject-specific noise ceiling by correlating their repeated measurements on a validation clip. Further, since we are only interested in the stimulus-driven response, we measure performance in voxels that exhibit high inter-subject correlations. We randomly split the 10 subjects into groups of 5, and correlate the mean activity of the two groups. We repeat this process 5 times and voxels that exhibit a mean correlation greater than 0.1 are identified as \textit{synchronous} voxels. We compute the mean normalized correlations across all synchronous voxels to achieve a single metric per subject, denoted as `Prediction accuracy'. We also correlate the predicted response of each subject against the predicted and true response of every other subject to obtain an $N \times N$ correlation matrix for shared models. 
To account for higher variability in measured versus predicted response, we normalize the rows and columns of this correlation matrix following \cite{pmid27124457}. 
\subsection{Demonstration of application:  personalized brain mapping}
To investigate if the proposed model is indeed capturing meaningful individual differences, we use the trained encoding model to predict fMRI activations for distinct visual object categories from the HCP task battery. 
Specifically, we predict brain response to visual stimuli (comprising faces, places, tools and body parts) from the HCP Working Memory (WM) task and use the $\textit{predicted}$ response to synthesize face and scene contrasts (FACES-AVG and PLACES-AVG respectively) for each individual. The predicted and true contrasts are thresholded to keep top $5\%$ of the voxels. 
%with highest activations in response to faces and scenes respectively versus the rest. 
We compute the Dice overlap between the predicted contrast for each subject against the true contrast of every subject (including self) to produce an $N \times N$ matrix for each contrast.

\section{Results}
\begin{figure}[!ht]
\includegraphics[width=0.9\textwidth]{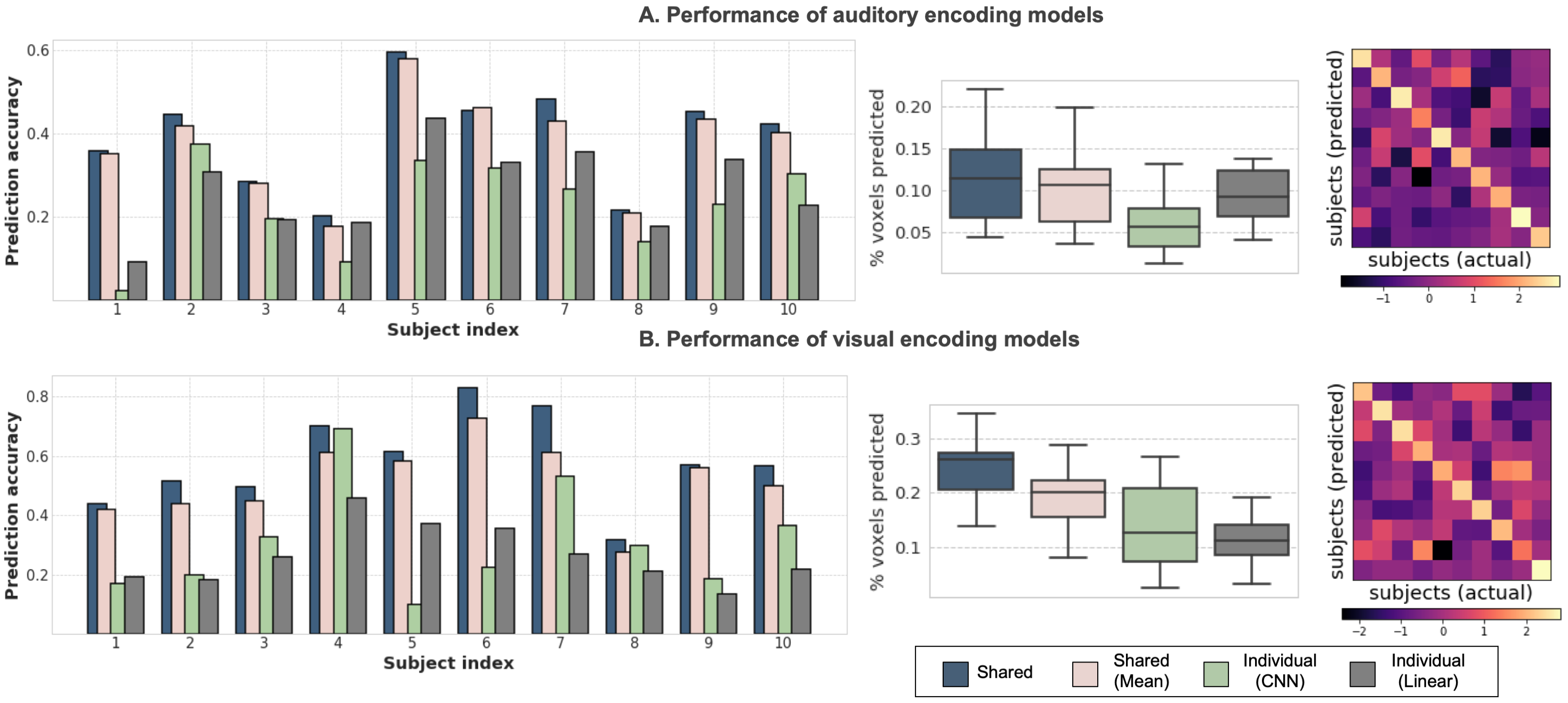}
\caption{\textbf{Quantitative evaluation}: Bar charts illustrate subject-wise prediction accuracy of all models, box plots depict the distribution over subjects for \% of synchronous voxels significantly predicted (p<0.05, FDR corrected). $N \times N$ correlation matrices depict the (normalized) correlation coefficient between predicted and measured responses.} \label{fig:quant}
\end{figure}

Figure \ref{fig:quant} shows prediction accuracy of the proposed (`Shared') and baseline methods for each subject. The performance improvement is striking between proposed and individual subject models, suggesting that a shared backbone architecture can significantly boost generalization. Comparative boxplots further show that the proposed method predicts a much higher percentage of the synchronous cortex than individual subject models. Further, the difference between `Shared' and `Shared (mean)' as well as the dominant diagonal structure in correlation matrices suggest that the proposed method is indeed capturing subject idiosyncrasies rather than predicting a group-averaged response. Further, while the CNN response model performs slightly better in visual encoding, it incurs a performance drop compared to linear regression in auditory encoding. This perhaps suggests that the boost in accuracy seen for shared models is largely due to inter-subject knowledge transfer rather than the convolutional response model itself.

\begin{figure}[!ht]
\begin{center}
\includegraphics[width=0.75\textwidth]{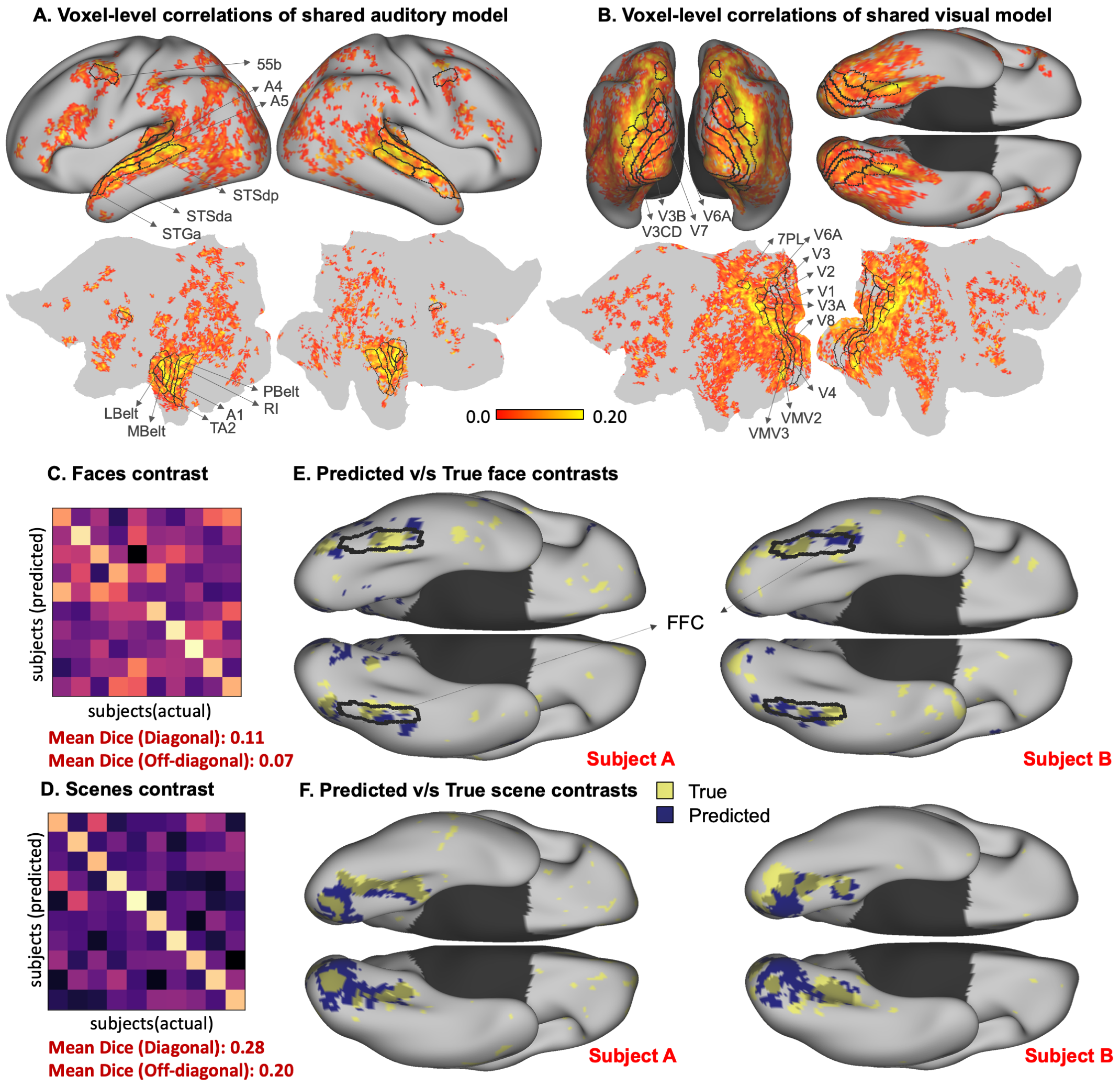}
\caption{(A), (B) Correlations between predicted response of the \textit{proposed} model and true time series of each voxel averaged across subjects. Only significantly predicted voxels are shown (p<0.05, FDR corrected). Dice matrices of predicted \textit{versus} true contrasts for (C) faces and (D) scenes stimuli. (E) \& (F) depict contrasts of two randomly selected subjects. ROIs are labelled from the HCP MMP parcellation~\cite{pmid27437579}.} 
\label{fig:comb}
\end{center}
\end{figure}

In Figure \ref{fig:comb}(A) \& \ref{fig:comb}(B), we visualize the un-normalized correlations between the predicted and measured fMRI response for the proposed models, averaged across subjects. For the auditory model, we see significant correlations in the parabelt auditory cortex, extending into the superior temporal sulcus and some other language areas (55b) as well. For the visual model, while we see significant correlations across the entire visual cortex (V1-V8), the performance is much better in higher-order visual regions, presumably because of the semantically rich features. The lower performance in early visual regions could also result from the dynamic nature of visual stimulation in movies.   
% \begin{figure}
% \begin{center}
% \includegraphics[width=0.9\textwidth]{Figures/vox_corrs.png}
% \caption{Correlations between predicted response of the \textit{proposed} model and true time series of each voxel averaged across subjects. Only significantly predicted voxels are shown (p<0.05, FDR corrected). ROIs are labelled from the HCP MMP parcellation} \label{fig:vox}
% \end{center}
% \end{figure}

% \begin{figure}
% \begin{center}

% \includegraphics[width=0.75\textwidth]{Figures/contrasts.png}
% \caption{Dice matrices of predicted \textit{versus} true contrasts for (A) faces and (B) scenes stimuli. (C) \& (D) depict contrasts of two randomly selected subjects.} \label{fig:contrast}
% \end{center}
% \end{figure}

Figure \ref{fig:comb}(C) \& \ref{fig:comb}(D) illustrate the ability of our proposed model to characterize individual differences even beyond the experimental paradigm it was trained on. The diagonal dominance in the dice matrix for both contrasts suggests that predicted contrasts are most similar to the same subject's true contrast. No prominent diagonal structure was observed for individual subject models, presumably because of their poor generalization to out-of-domain stimuli from the HCP task battery. Further, predicted contrasts consistently highlight known areas for face and scene processing, namely the fusiform face area\cite{pmid9151747} and parahippocampal areas\cite{pmid21957240} respectively.

\vspace{-10mm}
\section{Discussion}
In this paper, we presented a framework for utilizing multi-subject fMRI data to improve individual-level neural encoding. We showcased our approach on both auditory and visual stimuli and demonstrated consistent improvement over competing approaches. Our experiments further suggest that a single experiment (free-viewing of movies) can characterize a multitude of brain processes at once. This has important implications for brain mapping which traditionally relies on a battery of carefully-constructed stimuli administered within block-designs. Inter-subject variability in response patterns induced by the complexity of naturalistic viewing can facilitate the development of novel imaging-based biomarkers. Neural encoding models are not constrained to modeling the response to a limited set of experimental stimuli; their good generalization performance suggests that they can capture broad theories of cognitive processing. Accurate, individualized neural encoding models can thus bring us one step closer to achieving the goal of biomarker discovery.  

\section*{Acknowledgements}
This work was supported by NIH grants R01LM012719 (MS), R01AG053949 (MS), R21NS10463401 (AK), R01NS10264601A1 (AK), the NSF NeuroNex grant 1707312 (MS), the NSF CAREER 1748377 grant (MS) and Anna-Maria and Stephen Kellen Foundation Junior Faculty Fellowship (AK).
% Several extensions could be considered for this work. What is the finest resolution till which feature maps can be shared is an empirical question that needs further investigation. Longer narratives or movies further have an inherent temporal structure; models that can capture long-range relationships are likely more useful for encoding brain responses in higher-order regions. Exploiting this temporal structure can also emphasize individual differences in social processing\cite{pmid31770637}. 

%Further, the contribution of different layers to the prediction at each voxel can reveal interesting clues into the cortical processing hierarchy. At present, the model is not amenable to probe this effect.

%To achieve end-to-end optimization of the shared without, we proposed a convolutional response model. 

%%At the same time, as shown in Figure \ref{fig:contrast}, they are able to capture individual differences in processing of these categories as well. 

%\paragraph{Sample Heading (Fourth Level)}

%\clearpage

\bibliographystyle{unsrt}
\bibliography{shared_encoding}

\begin{thebibliography}{10}

\bibitem{pmid15016991}
U.~Hasson, Y.~Nir, I.~Levy, G.~Fuhrmann, and R.~Malach.
\newblock {{I}ntersubject synchronization of cortical activity during natural
  vision}.
\newblock {\em Science}, 303(5664):1634--1640, Mar 2004.

\bibitem{pmid31257145}
S.~Sonkusare, M.~Breakspear, and C.~Guo.
\newblock {{N}aturalistic {S}timuli in {N}euroscience: {C}ritically
  {A}cclaimed}.
\newblock {\em Trends Cogn. Sci. (Regul. Ed.)}, 23(8):699--714, Aug 2019.

\bibitem{Schultz2009}
J.~Schultz and K.~S. Pilz.
\newblock {{N}atural facial motion enhances cortical responses to faces}.
\newblock {\em Exp Brain Res}, 194(3):465--475, Apr 2009.

\bibitem{hcp}
M.~F. Glasser, S.~N. Sotiropoulos, J.~A. Wilson, T.~S. Coalson, B.~Fischl,
  J.~L. Andersson, J.~Xu, S.~Jbabdi, M.~Webster, J.~R. Polimeni, D.~C.
  Van~Essen, and M.~Jenkinson.
\newblock {{T}he minimal preprocessing pipelines for the {H}uman {C}onnectome
  {P}roject}.
\newblock {\em Neuroimage}, 80:105--124, Oct 2013.

\bibitem{pmid20004608}
U.~Hasson, R.~Malach, and D.~J. Heeger.
\newblock {{R}eliability of cortical activity during natural stimulation}.
\newblock {\em Trends Cogn. Sci. (Regul. Ed.)}, 14(1):40--48, Jan 2010.

\bibitem{Chen2015ARF}
Po-Hsuan~Cameron Chen, Janice Chen, Yaara Yeshurun, Uri Hasson, James~V. Haxby,
  and Peter~J. Ramadge.
\newblock A reduced-dimension {fMRI} shared response model.
\newblock In {\em NIPS}, 2015.

\bibitem{pmid30513462}
G.~Varoquaux and R.~A. Poldrack.
\newblock {{P}redictive models avoid excessive reductionism in cognitive
  neuroimaging}.
\newblock {\em Curr. Opin. Neurobiol.}, 55:1--6, 04 2019.

\bibitem{Kell2018a}
Alexander~J.E. Kell, Daniel~L.K. Yamins, Erica~N. Shook, Sam~V.
  Norman-Haignere, and Josh~H. McDermott.
\newblock {A Task-Optimized Neural Network Replicates Human Auditory Behavior,
  Predicts Brain Responses, and Reveals a Cortical Processing Hierarchy}.
\newblock {\em Neuron}, 98(3):630--644.e16, may 2018.

\bibitem{pmid26157000}
U.~Guclu and M.~A. van Gerven.
\newblock {{D}eep {N}eural {N}etworks {R}eveal a {G}radient in the {C}omplexity
  of {N}eural {R}epresentations across the {V}entral {S}tream}.
\newblock {\em J. Neurosci.}, 35(27):10005--10014, Jul 2015.

\bibitem{pmid24812127}
D.~L. Yamins, H.~Hong, C.~F. Cadieu, E.~A. Solomon, D.~Seibert, and J.~J.
  DiCarlo.
\newblock {{P}erformance-optimized hierarchical models predict neural responses
  in higher visual cortex}.
\newblock {\em Proc. Natl. Acad. Sci. U.S.A.}, 111(23):8619--8624, Jun 2014.

\bibitem{Wen2018a}
Haiguang Wen, Junxing Shi, Yizhen Zhang, Kun~Han Lu, Jiayue Cao, and Zhongming
  Liu.
\newblock {Neural encoding and decoding with deep learning for dynamic natural
  vision}.
\newblock {\em Cerebral Cortex}, 28(12):4136--4160, dec 2018.

\bibitem{pmid27138646}
J.~Dubois and R.~Adolphs.
\newblock {{B}uilding a {S}cience of {I}ndividual {D}ifferences from
  f{M}{R}{I}}.
\newblock {\em Trends Cogn. Sci. (Regul. Ed.)}, 20(6):425--443, 06 2016.

\bibitem{Wen2018b}
Haiguang Wen, Junxing Shi, Wei Chen, and Zhongming Liu.
\newblock {Transferring and generalizing deep-learning-based neural encoding
  models across subjects}.
\newblock {\em NeuroImage}, 176:152--163, aug 2018.

\bibitem{Guclu2017}
Umut G{\"{u}}{\c{c}}l{\"{u}} and Marcel~A.J. van Gerven.
\newblock {Increasingly complex representations of natural movies across the
  dorsal stream are shared between subjects}.
\newblock {\em NeuroImage}, 145:329--336, jan 2017.

\bibitem{Lin2016FeaturePN}
Tsung-Yi Lin, Piotr Doll{\'a}r, Ross~B. Girshick, Kaiming He, Bharath
  Hariharan, and Serge~J. Belongie.
\newblock Feature pyramid networks for object detection.
\newblock {\em 2017 IEEE Conference on Computer Vision and Pattern Recognition
  (CVPR)}, pages 936--944, 2016.

\bibitem{He2015DeepRL}
Kaiming He, Xiangyu Zhang, Shaoqing Ren, and Jian Sun.
\newblock Deep residual learning for image recognition.
\newblock {\em 2016 IEEE Conference on Computer Vision and Pattern Recognition
  (CVPR)}, pages 770--778, 2015.

\bibitem{Hershey2016CNNAF}
Shawn Hershey, Sourish Chaudhuri, Daniel P.~W. Ellis, Jort~F. Gemmeke, Aren
  Jansen, R.~Channing Moore, Manoj Plakal, Devin Platt, Rif~A. Saurous, Bryan
  Seybold, Malcolm Slaney, Ron~J. Weiss, and Kevin~W. Wilson.
\newblock Cnn architectures for large-scale audio classification.
\newblock {\em 2017 IEEE International Conference on Acoustics, Speech and
  Signal Processing (ICASSP)}, pages 131--135, 2016.

\bibitem{Deng2009ImageNetAL}
Jia Deng, Wei Dong, Richard Socher, Li-Jia Li, Kai Li, and Fei-Fei Li.
\newblock Imagenet: A large-scale hierarchical image database.
\newblock {\em 2009 IEEE Conference on Computer Vision and Pattern
  Recognition}, pages 248--255, 2009.

\bibitem{AbuElHaija2016YouTube8MAL}
Sami Abu-El-Haija, Nisarg Kothari, Joonseok Lee, Apostol Natsev, George
  Toderici, Balakrishnan Varadarajan, and Sudheendra Vijayanarasimhan.
\newblock Youtube-8m: A large-scale video classification benchmark.
\newblock {\em ArXiv}, abs/1609.08675, 2016.

\bibitem{Detectron2018}
Ross Girshick, Ilija Radosavovic, Georgia Gkioxari, Piotr Doll\'{a}r, and
  Kaiming He.
\newblock Detectron.
\newblock \url{https://github.com/facebookresearch/detectron}, 2018.

\bibitem{vggish}
S.~Hershley and et. al.
\newblock Models for audioset: A large scale dataset of audio events.
\newblock
  \url{https://github.com/tensorflow/models/tree/master/research/audioset/vggish},
  2016.

\bibitem{Kingma2014AdamAM}
Diederik~P. Kingma and Jimmy Ba.
\newblock Adam: A method for stochastic optimization.
\newblock {\em CoRR}, abs/1412.6980, 2014.

\bibitem{hcp2}
D.~C. Van~Essen, K.~Ugurbil, E.~Auerbach, D.~Barch, T.~E. Behrens, R.~Bucholz,
  A.~Chang, L.~Chen, M.~Corbetta, S.~W. Curtiss, S.~Della~Penna, D.~Feinberg,
  M.~F. Glasser, N.~Harel, A.~C. Heath, L.~Larson-Prior, D.~Marcus,
  G.~Michalareas, S.~Moeller, R.~Oostenveld, S.~E. Petersen, F.~Prior, B.~L.
  Schlaggar, S.~M. Smith, A.~Z. Snyder, J.~Xu, and E.~Yacoub.
\newblock {{T}he {H}uman {C}onnectome {P}roject: a data acquisition
  perspective}.
\newblock {\em Neuroimage}, 62(4):2222--2231, Oct 2012.

\bibitem{pmid27894889}
A.~T~Vu, K.~Jamison, M.~F. Glasser, S.~M. Smith, T.~Coalson, S.~Moeller, E.~J.
  Auerbach, K.~Ugurbil, and E.~Yacoub.
\newblock {{T}radeoffs in pushing the spatial resolution of f{M}{R}{I} for the
  7{T} {H}uman {C}onnectome {P}roject}.
\newblock {\em Neuroimage}, 154:23--32, 07 2017.

\bibitem{pmid27124457}
I.~Tavor, O.~Parker~Jones, R.~B. Mars, S.~M. Smith, T.~E. Behrens, and
  S.~Jbabdi.
\newblock {{T}ask-free {M}{R}{I} predicts individual differences in brain
  activity during task performance}.
\newblock {\em Science}, 352(6282):216--220, Apr 2016.

\bibitem{pmid9151747}
N.~Kanwisher, J.~McDermott, and M.~M. Chun.
\newblock {{T}he fusiform face area: a module in human extrastriate cortex
  specialized for face perception}.
\newblock {\em J. Neurosci.}, 17(11):4302--4311, Jun 1997.

\bibitem{pmid21957240}
S.~Nasr, N.~Liu, K.~J. Devaney, X.~Yue, R.~Rajimehr, L.~G. Ungerleider, and
  R.~B. Tootell.
\newblock {{S}cene-selective cortical regions in human and nonhuman primates}.
\newblock {\em J. Neurosci.}, 31(39):13771--13785, Sep 2011.

\bibitem{pmid27437579}
M.~F. Glasser, T.~S. Coalson, E.~C. Robinson, C.~D. Hacker, J.~Harwell,
  E.~Yacoub, K.~Ugurbil, J.~Andersson, C.~F. Beckmann, M.~Jenkinson, S.~M.
  Smith, and D.~C. Van~Essen.
\newblock {{A} multi-modal parcellation of human cerebral cortex}.
\newblock {\em Nature}, 536(7615):171--178, 08 2016.

\end{thebibliography}

\end{document}